\documentclass[conference]{IEEEtran}
\IEEEoverridecommandlockouts
\usepackage{cite}
\usepackage{amsmath,amssymb,amsfonts}
\usepackage{braket}
\usepackage{algorithmic}
\usepackage{graphicx}
\usepackage{textcomp}
\usepackage{hyperref}
\usepackage[shortlabels]{enumitem}
\usepackage{pifont}
\usepackage{amsthm}
\usepackage[dvipsnames]{xcolor}
\usepackage{multirow, tabularx}
\usepackage{cancel}
\usepackage{booktabs}

\usepackage{graphicx} 

\usepackage{orcidlink}

\usepackage{caption} 
\definecolor{purple}{rgb}{0.75, 0.00, 1.00} 

\definecolor{lukas}{HTML}{76c11f}

\usepackage{listingsutf8}
\definecolor{codegreen}{rgb}{0,0.6,0}
\definecolor{codegray}{rgb}{0.5,0.5,0.5}
\definecolor{codepurple}{rgb}{0.58,0,0.82}
\definecolor{backcolour}{rgb}{0.95, 0.95, 0.95}

\lstdefinestyle{codeblock}{
  backgroundcolor=\color{backcolour},
  commentstyle=\color{codegreen},
  keywordstyle=\color{blue},
  numberstyle=\tiny\color{codegray},
  stringstyle=\color{codepurple},
  basicstyle=\footnotesize,
  escapechar=\¢, 
  otherkeywords={with},
  breakatwhitespace=false,
  breaklines=true,
  captionpos=b,
  keepspaces=true,
  language=Python,
  numbers=left,
  numbersep=2pt,
  showspaces=false,
  showstringspaces=false,
  showtabs=false,
  tabsize=2,
  basicstyle=\ttfamily\footnotesize,
  inputencoding=utf8,
  upquote=true,
}
\newtheorem{definition}{Definition}
\newcommand{\cmark}{\ding{51}}
\newcommand{\xmark}{\ding{55}}

\makeatletter
\newcommand{\linebreakand}{%
  \end{@IEEEauthorhalign}
  \hfill\mbox{}\par
  \mbox{}\hfill\begin{@IEEEauthorhalign}
}
\makeatother

\title{Why Are We Unrolling? The Importance of Structured Quantum Programs for Compilation}

\author{
\IEEEauthorblockN{
Damian Rovara\IEEEauthorrefmark{1}\IEEEauthorrefmark{2}\orcidlink{0009-0000-7416-9149}, 
Daniel Haag\IEEEauthorrefmark{1}\IEEEauthorrefmark{6}\orcidlink{0000-0001-5069-6884}, 
Mark Koch\IEEEauthorrefmark{3}\IEEEauthorrefmark{4}\orcidlink{0000-0001-8511-2703},
Josh Izaac\IEEEauthorrefmark{5}\orcidlink{0000-0003-2640-0734},
Seyon Sivarajah\IEEEauthorrefmark{4}\orcidlink{0000-0002-7332-5485},
Robert Wille\IEEEauthorrefmark{2}\IEEEauthorrefmark{6}\orcidlink{0000-0002-4993-7860},\\
Agust\'in Borgna\IEEEauthorrefmark{4}\orcidlink{0000-0002-1688-1370},
Lukas Burgholzer\IEEEauthorrefmark{2}\IEEEauthorrefmark{6}\orcidlink{0000-0003-4699-1316},
Brad Chase\IEEEauthorrefmark{7}\orcidlink{0000-0002-7696-8179},
Olivia Di Matteo\IEEEauthorrefmark{8}\orcidlink{0000-0002-1372-7706},
David Ittah\IEEEauthorrefmark{5}\orcidlink{0000-0003-0975-6448}\\
}

\IEEEauthorblockA{\IEEEauthorrefmark{1}\textit{Technical University of Munich}, Munich, Germany}

\IEEEauthorblockA{\IEEEauthorrefmark{2} Corresponding author: 
damian.rovara@tum.de}

\IEEEauthorblockA{\IEEEauthorrefmark{3}\textit{University of Oxford}, Oxford, UK}

\IEEEauthorblockA{\IEEEauthorrefmark{4}\textit{Quantinuum}, Cambridge, United Kingdom}

\IEEEauthorblockA{\IEEEauthorrefmark{5}\textit{Xanadu}, Toronto, ON, M5G 2C8, Canada}

\IEEEauthorblockA{\IEEEauthorrefmark{6}\textit{MQSC}, Garching near Munich, Germany}

\IEEEauthorblockA{\IEEEauthorrefmark{7}\textit{Unitary Foundation}, San Francisco, CA, USA}

\IEEEauthorblockA{\IEEEauthorrefmark{8}
\textit{The University of British Columbia}, Vancouver BC, Canada}
}

\begin{document}

\maketitle

\begin{abstract}
As quantum software stacks scale up to support future fault-tolerant quantum hardware and algorithms, quantum compilation is becoming an increasingly important component of the stack. How do we ensure that our software stacks support dynamic algorithms---including patterns such as mid-circuit measurement feedforward and repeat-until-success---with hundreds of logical qubits and billions of quantum operations? To do so, we must re-think how we represent quantum programs beyond straight-line circuits, to representations that include classical structure and dynamism, and make this the default representation to consider when performing quantum compilation. In this work, we present important patterns and algorithms from fault-tolerant quantum applications which admit a structured representation that we argue is crucial to preserve, and set a challenge to the community to compile such representations without unrolling them into straight-line quantum circuits. We also explore the status quo of structured program support in quantum software, and ask ourselves the rhetorical question: how much more efficient can we make quantum compilation tooling when we take into account the additional information from classical structure?
\end{abstract}


\section{Introduction}

The accelerating pace of quantum hardware towards fault-tolerance necessitates software capable of orchestrating the execution of quantum algorithms at scale. 
Resource estimates for solving real-world problems with active error correction often yield physical qubit counts on the order of millions and gate counts exceeding billions \cite{beverland2022assessingrequirementsscalepractical, agrawal2024quantifyingfaulttolerantsimulation,nguyen2025quantumcomputingcorrosionresistantmaterials}. 
Managing such a large amount of operations is a complex scheduling problem;
determining that sequence of operations in the first place is a separate challenge addressed by \emph{quantum compilation}.

Quantum compilation translates an implementation of a quantum algorithm into executable instructions. 
Effective quantum compilers enable users to implement algorithms with minimal concern for execution details such as error correction, gates on individual qubits, or processor architecture.
Historically, quantum algorithms were primarily expressed as explicit logical circuits. Thus, ``quantum compilation" was synonymous with quantum circuit synthesis and optimization, which unrolls a fully-specified circuit into a flat list of gates (or graph) and applies a sequence of transformations. These transformations are usually equivalence-preserving and may include circuit synthesis and decomposition, rewriting and pattern-matching optimization, transpilation into different gate sets, and modifications based on hardware-specific constraints.

Significant recent effort has been invested in programming frameworks that raise the abstraction level beyond the circuit model \cite{qiskit, bergholm2018pennylane, seidel2024qrisp, cirq, cross2022, svore2018qsharp, bichsel2020silq, koch2025},
allowing algorithms to be expressed using familiar constructs like subroutines, loops, and classical control flow.
This raises the question of how to compile and optimize such programs. Hardware execution involves streaming a list of instructions to a device.
The list can be generated in advance, by unrolling loops or creating multiple program versions to account for branching, and optimized with circuit transformations.
However, this has scalability implications as many circuit transformations require solving computationally hard problems (e.g., qubit allocation is NP-complete \cite{siraichi2018qubit} and qubit routing is NP-hard \cite{ito2025algorithmic}); incorporating error correction adds yet another layer of complexity.
Moreover, any change in program parameters incurs recompilation.

Parametric compilation is one approach to avoid recompilation \cite{karalekas2020parametric}. In parametric compilation, programs are compiled with symbolic input parameters, whose values are provided at execution time --- avoiding the need for re-compilation with new parameter values.
Python-based frameworks such as JAX~\cite{jax2018github}, Catalyst~\cite{Ittah2024}, and Qrisp~\cite{seidel2024qrisp} often implement parametric compilation via what's known as just-in-time (JIT) compilation. In these frameworks, the first time a program is executed in the Python interpreter, tracers are passed through a program instead of concrete input values, which then record the program's instruction stream (possibly including structure). This representation is compiled and cached for reuse in subsequent executions, avoiding recompilation provided the type signature of the inputs doesn’t change.

\begin{figure}
    \centering
    \includegraphics[width=0.45\textwidth]{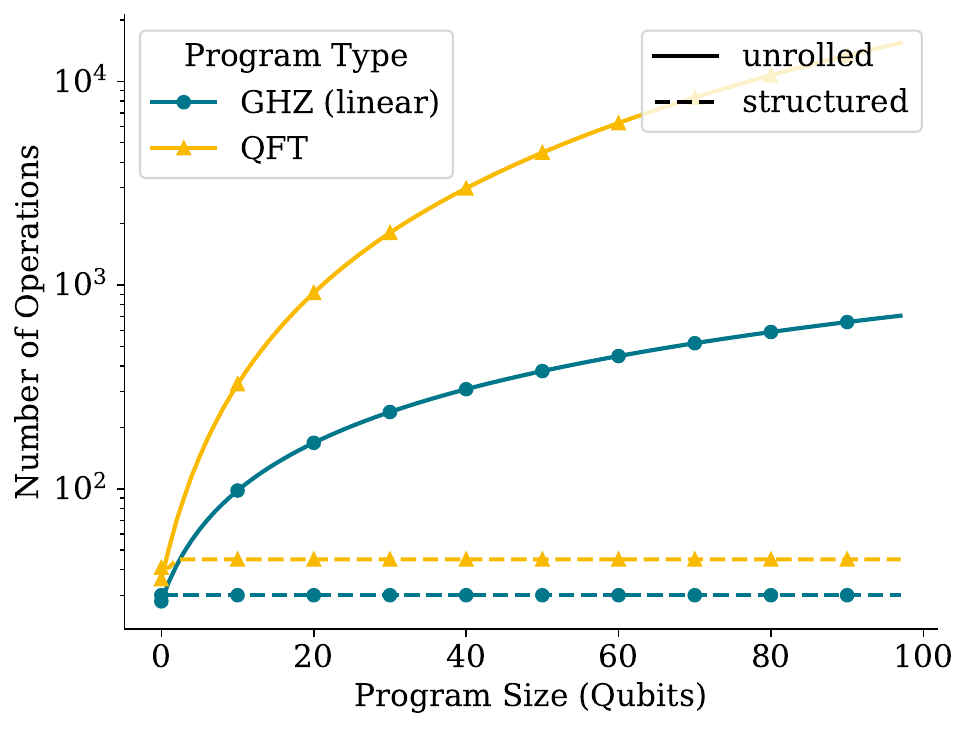}
    \caption{Preserving program structure can enable constant compilation times, even when the problem instance size increases. On the other hand, unrolling loops generally leads to polynomial or even exponentially-increasing program size. Programs used for this comparison were generated using \texttt{mqt-cc}~\cite{burgholzer2026a}. See \autoref{sec:challenge} for details on benchmark~programs.
    }
    \label{fig:unrolling}
\end{figure}

However, avoiding recompilation is not the only challenge. Some quantum algorithms fundamentally \emph{cannot} be expressed as one flat circuit. This includes: repeat-until-success patterns in circuit synthesis \cite{Paetznick2013RUS}, block encodings for Hamiltonian simulation \cite{Low2017QSP, Low2019hamiltonian} and quantum singular value transformations \cite{gilyen2019qsvt}; magic state distillation \cite{bravyi2005msd}; active error correction \cite{Campbell2017ftqc}; and Shor's algorithm \cite{shor1997}.
Some intermediate representations, enabled by classical compiler infrastructure, can preserve and exploit structure in these algorithms \cite{Ittah2024, seidel2024qrisp, qir, cross2022, burgholzer2026a, jeff, mlir, llvm},
leading to compilation times and program sizes independent of problem size.
Such constant scaling, depicted graphically for sample problems in \autoref{fig:unrolling}, was demonstrated empirically in multiple frameworks \cite{ittah2022,ittah2025constanttimehybridcompilationshors, seidel2024qrisp}.

A major roadblock is that, unless unrolled, programs with structure are incompatible with compilers that only support flat circuit transformations. Novel, dedicated tooling is required to both support and exploit such structure during optimization.
While quantum circuit optimization is well-studied, there are fewer approaches to optimization of structured quantum programs. 
Often, such optimizations are the result of human intuition and manual implementation \cite{ittah2025constanttimehybridcompilationshors}.
Existing systematic methods rely on underlying mathematical structures that limit their utility. For example, assertion-based optimization uses the Z3 theorem prover to establish pre- and post-conditions that enable removal of trivial gates and subroutines \cite{haener2020assertion}. Relational analysis of program input and output variables can be used to formulate techniques such as phase folding and perform optimizations around loops, but encounters challenges with gates that create superpositions \cite{amy2025linear}. 

The tension between preserving program structure and flattening it into a linear instruction stream is not unique to quantum computing; it has shaped classical accelerator compilation for decades.
Polyhedral compilers such as \emph{Pluto} and \emph{PPCG}~\cite{bondhugula2008pluto, verdoolaege2013polyhedral} retain loop nests as an explicit iteration space instead of unrolling them.
The same principle applies to intermediate representations (IRs) built specifically for accelerator generation: recent designs deliberately keep loops and conditionals in the IR rather than eliminating them up front to enable further optimizations---lowering control flow only as a final step~\cite{nigam2021compiler, kim2024unifying}.
Database query engines are a further classical instance: a query is optimized as a relational-algebra plan and only then compiled to machine code, so that the structural information in the plan is exploited before any per-tuple work is materialized~\cite{neumann2011efficiently}.
In contrast, reconstructing loops from already flattened code is an active field of research, aiming to reduce code size and recover analysis precision~\cite{grosser2015optimistic, rocha2022loop}.
Where flattening does occur, it is a local and deliberate choice---unrolling a small inner loop for throughput---not a generic expansion of the program~\cite{davidson1996aggressive}.

In light of today's landscape, we argue that preserving program structure is not just optional, but \emph{essential} for scalable quantum compilation. This necessitates research on both the representation and optimization of structured programs. In this work, we make three contributions towards that goal: 
\begin{itemize}
  \item We provide a formal definition of \emph{structured quantum programs} and their constituent elements.
  \item We survey the software landscape and assess current frameworks' level of support for structured programs 
  \item We propose a diverse set of programs, with varying amounts of structural elements, to serve as a benchmark set for current and future compilers. We highlight explicit examples where preserving structure in the representation is essential for compilation.
\end{itemize}

\section{Structured Quantum Programs}

\subsection{Definition}
\label{sec:definition}


\begin{table*}[htbp]
\caption{Elements of structured quantum programs. The Tier, described in  \autoref{sec:taxonomy}, differentiates between elements that are merely advantageous (Tier 1) vs. required (Tier 2) for a complete, universal program representation. Note that conditioning and iteration can appear together in a single program construct, like a while loop.
  }
  \label{tab:structural_elements}
  \centering
  \begin{tabular}{|c|p{4.2cm}|c|p{9.4cm}|}
  \hline 
  \textbf{Tier} &
  \textbf{Primitives of structured programs} & \textbf{Description} \\  \hline
1 & StatIter    & Program uses iteration (loops) with a constant number of repetitions.\\
2 & DynIter    & Iteration (loop) termination conditions depend on values computed at runtime\\
1 & DynVal   & Gates use classical values computed at runtime (e.g., rotation angles from arrays)\\
1 & CCond    & Conditional depends on classical values that are not results of quantum measurements\\
2 & QCond    & Conditionals depend on results of one or more quantum measurements\\
1 & DynAlloc & Qubits are allocated at runtime (e.g., inside loop bodies)\\
2 & DynIdx   & Gates are applied to qubits with non-constant indices (e.g., loop variable)\\ \hline
  \end{tabular}
  
\end{table*}

A \emph{quantum program} is a computer program wherein one or more instructions are executed on a quantum device. A characteristic feature of \emph{structured} quantum programs is a representation with hierarchical structure beyond a flat list of instructions, such as subroutines or structured control flow. 
More concretely, structured program execution does not proceed in a strictly sequential manner, as formalized below.

\begin{definition}
Let $P$ be a quantum program, and $c_P$ an integer-valued program counter indicating the line of $P$ currently being executed. $P$ is a \emph{structured quantum program} if $c_P$ does not evolve exclusively by unit increments, i.e., there exist execution steps where $c_P \not\mapsto c_P + 1$.
\label{def:structured-programs}
\end{definition}

\setlength{\tabcolsep}{10pt} 
\begin{table*}[htbp]
\caption{Examples of static and dynamic (un)structured quantum programs using OpenQASM 2 \cite{cross2017openqasm2} and OpenQASM 3 \cite{cross2022} syntax. In both unstructured programs, the program counter increases linearly. In the structured static program, the counter jumps back when the \texttt{ghz\_cnots} subroutine executes, then follows loop execution. In the structured dynamic program, the counter cycles through the \texttt{while} loop; note such dynamically-bounded \texttt{while} loops are not supported in OpenQASM 2.}
    \label{tab:structured-program-matrix}
\centering
    \begin{tabular}
    {| p{3.7cm} | p{3.7cm}| p{3.7cm} |p{3.8cm}|}
    \hline
   \multicolumn{2}{|c|}{\textbf{Static}} & \multicolumn{2}{|c|}{\textbf{Dynamic}}  \\ \hline
    \textbf{Unstructured} & \textbf{Structured} & \textbf{Unstructured} & \textbf{Structured} \\ \hline
\begin{lstlisting}
qreg q[4];
h q[0];
cx q[0], q[1];
cx q[0], q[2];
cx q[0], q[3];
\end{lstlisting} &
\begin{lstlisting}
def ghz_cnots(q) {
    for i in [1:3] {
       cx q[0], q[i];
    }
}
qreg q[4];
h q[0];
ghz_cnots(q);
\end{lstlisting}
& \begin{lstlisting}
qreg q[2];
creg c[1];
ry(0.1) q[0];
measure q[0] -> c;
if (c == 1) x q[1];
\end{lstlisting} &
\begin{lstlisting}
qreg q[2];
creg c[1];
while (c != 1) {
    ry(0.1) q[0];
    measure q[0] -> c;
    reset q[0];
}
x q[1];
\end{lstlisting} \\ \hline
    \end{tabular}
\end{table*}
\setlength{\tabcolsep}{6pt}

Seven key primitives of structured programs are outlined in \autoref{tab:structural_elements}, and examples of equivalent structured and non-structured programs are presented in \autoref{tab:structured-program-matrix}. 
Program types are further subdivided based on data dependence.
If the quantum instructions in $P$ are fully determined before execution, $P$ is a \emph{static} quantum program. In \emph{dynamic} quantum programs, $c_P$, instructions at $c_P$, or instruction parameters depend on classical values or quantum measurements resolved at runtime.

Many primitives from \autoref{tab:structural_elements} occur naturally in quantum algorithms, though their use may be optional. A concrete example is included below in a pseudo-Pythonic implementation of Shor's algorithm, based on the implementation in Ref. \cite{ittah2025constanttimehybridcompilationshors}. Instances of each primitive are noted in the comments.

\begin{lstlisting}
def shors_algorithm(N):
  a = randint(2, N - 1)

  # CCond: conditional on dynamic classical value
  p = gcd(a, N)
  if p != 1:
    return p, N // p
    
  # DynAlloc: qubit register size based on bitwidth 
  n_bits = floor(log2(N)) + 1
  est_q = qalloc(1)
  target_q = qalloc(n_bits)
  aux_q = qalloc(n_bits + 2)

  meas = zeros((n_bits, ))
  cumul_phase = 0.0
  idx = 0
  pow_a = a

  def run_qpe():
    PauliX(target_q[-1])

    # DynIter,CCond: loop termination based on n_bits, a
    while idx < n_bits and pow_a != 1:
      Hadamard(est_q)

      # DynIdx: qubit indices in loops based on n_bits
      # DynVal: gates will depend on binary form of a
      ctrl_ua(N, pow_a, est_q, target_q, aux_q)

      PhaseShift(cumul_phase, est_q)
      Hadamard(est_q)

      meas[idx] = measure(est_q, reset=True)

      # QCond: conditional on measurement result
      if meas[idx] == 1:
        cumul_phase += -2 * pi / 2 ** (idx + 1)
      
      idx += 1
      pow_a = a ** (2**idx) % N
    return meas

  p, q = 0, 0

  # StatIter: statically-bounded loop
  for _ in range(100):
    sample = run_qpe()
    guess_r = samples_to_order(sample, N)

    if guess_r % 2 == 0:
        # ... 

  return p, q
\end{lstlisting}

This is a dynamic structured quantum program: a random value \texttt{a}, selected at runtime, affects classical conditionals, loop termination conditions, and quantum gates. Array and qubit register sizes depend on the bit width of $N$, and there are conditionals based on measurement results.
Quantum gates are executed in multiple program regions. Most are in  the \texttt{while} region's modular exponentiation subroutine (\texttt{ctrl\_ua}), wherein gates, loops, and register indexing are based on \texttt{a} and $N$.
Optimizations using the runtime value of \texttt{a} can also be incorporated (e.g., removing controlled operations based on its binary representation, as done in \cite{ittah2025constanttimehybridcompilationshors}).

In Shor's algorithm, the circuit depends intimately on $a$ and $N$ so it will be different each time the algorithm executes.
Preserving structure where possible allows compilation of a more compact representation, which is more scalable than immediately unrolling.
Of course, some algorithms may eventually be unrolled prior to hardware execution.
However, the key is that sometimes structural elements are essential: the dynamic loop boundary (line 24) means the algorithm \emph{cannot} be represented by a flat circuit before execution.
In what follows, we provide a more comprehensive exploration of how the structural elements in \autoref{tab:structural_elements} impact compilation.



\section[Compiling with Structure]{Compiling with Structure:\\ Goals and Requirements}
This section discusses the core aspects of compiling structured quantum programs, dividing structural elements into a system of two tiers and discussing their requirements.

\subsection{A Taxonomy of Structured Quantum Program Features}
\label{sec:taxonomy}

To address the two core concerns of computational bottlenecks and expressivity, we revisit the structural elements defined in \autoref{tab:structural_elements}.
Rather than grouping elements by syntactic similarity, we categorize them by their impact on the quantum compilation pipeline.
In doing so, we create a taxonomy of structural elements consisting of a baseline and two distinct tiers: those enabling more efficient compilation, and those strictly required for universal execution of input programs.

\subsection*{Tier 0: \emph{Straight-Line Quantum Circuits}}
As a baseline, we define Tier 0 features to represent standard quantum circuits devoid of high-level control flow, i.e., purely sequential application of quantum gates and measurements.
While this representation is sufficient for many near-term algorithms, it fails to capture the complexity and scalability required by fault-tolerant quantum circuits or advanced algorithms without resorting to massive, unrolled circuits.

\subsection*{Tier 1: \emph{Structure as an Enabler}}
This tier of structural elements comprises features that are not a strict necessity for purely expressing a quantum program, but provide significant advantages by enabling advanced optimizations, reducing code size, and encoding detailed semantic intent.
The defining aspect of this tier is that its structural elements can, in principle, be deterministically translated into straight-line circuit primitives at compile time (e.g., by fully unrolling statically-bounded loops).
Consequently, existing quantum circuit compilers can achieve a baseline level of compatibility simply by implementing such translations as preprocessing steps.
However, to truly benefit from this tier, compilers must evolve beyond simple unrolling and preserve these elements natively in their IRs.
This tier includes the following elements and their~benefits:
\begin{itemize}
    \item \emph{Statically-bounded loops} reduce redundancies in the IR.
    \item \emph{Conditionals on classical values} allow the conditional application of quantum operations in a way that can be analyzed statically.
    \item \emph{Dynamic classical values} allow parametrized operations to be given in a more general representation.
    
    \item \emph{Dynamic qubit allocation} allows qubits to be confined to the parts of the code where they are required.
\end{itemize}

\subsection*{Tier 2: \emph{Structure as a Requirement}}
In contrast to the enabling nature of Tier 1, this tier consists of structural elements that are strictly required for the universal execution of complex quantum programs.
The defining characteristic of these elements is that they inherently depend on dynamic or runtime information.
Consequently, they cannot be preprocessed or unrolled into straight-line circuits under any circumstances.
Compilers must explicitly support and handle these features to even be compatible with the programs that employ them.
This tier includes the following elements, as well as the reasons why they cannot be reduced to straight-line circuits:
\begin{itemize}
    \item \emph{Dynamically-bounded loops} cannot be unrolled at compile time as iteration counts depend on runtime values and cannot be computed in advance.
    \item \emph{Dynamic qubit indexing} obscures data dependencies and cannot be replaced by static qubit indexing.
    \item \emph{Conditionals on measurement results} limit static analysis of conditional branches and loops and cannot be expressed by static circuit primitives.
\end{itemize}

\subsection{A Gap in Representation}

To prepare for the rising complexity of quantum programs, a quantum compilation framework must be able to address both tiers of structural elements presented above.
Unfortunately, the current landscape of quantum software is heavily fragmented, with varying levels of support for these critical elements.
Modern quantum execution relies on a multi-stage compilation pipeline, ranging from high-level front-end programming languages down through several IRs to formulate the target instructions for the quantum hardware.
In this pipeline, the actual support for structured quantum programs is ultimately dictated by the ``lowest common denominator'' among all involved layers.
It is simply not enough for only certain stages or intermediate formats to be capable of expressing these features.
To fully exploit structured programs, these elements need to be broadly supported across the entire toolchain.
More importantly, compilers must actively be able to transform their IR while preserving these features so that optimization passes can leverage them directly.

\newcommand{\cdouble}{\checkmark\!\!\checkmark}
\newcommand{\cyes}{\checkmark}
\newcommand{\cno}{\xmark}

\begin{table}
\centering
\caption{Comparison of common compilation frameworks and their ability to handle structural primitives. 
}
\label{tab:representation-compilers-comp}
\scalebox{0.94}{
\begin{tabular}{|c|c|c|c|c|c|c|c|c|} 
\hline
\rule{0pt}{4.8ex}
Framework & StatIter    & DynIter    & \shortstack[c]{Dyn\\Val} & CCond    & QCond    & \shortstack[c]{Dyn\\Alloc} & \shortstack[c]{Dyn\\Idx} \\\hline
Qiskit    & \cyes    & \cno     & \cyes                    & \cyes    & \cyes    & \cyes                      & \cno                     \\\hline
TKETv1    & \cno     & \cno     & \cno                     & \cyes    & \cyes    & \cno                       & \cno                     \\\hline
TKETv2    & \cdouble & \cdouble & \cdouble                 & \cdouble & \cdouble & \cyes                      & \cdouble                 \\\hline
Catalyst  & \cdouble & \cdouble & \cdouble                 & \cdouble & \cdouble & \cdouble                   & \cdouble                 \\\hline
mqt-cc    & \cyes    & \cyes    & \cdouble                 & \cdouble & \cdouble & \cdouble                   & \cyes                    \\\hline
\end{tabular}
}
\\
\begin{center}
    \footnotesize \hspace{0.5cm} \cno: incompatible \hfill \cyes: compatible \hfill \cdouble: exploitable \hspace{1cm}
\end{center}
\vspace{-1em}
\end{table}

To assess the current state of this software pipeline, \autoref{tab:representation-compilers-comp} evaluates the support for structural primitives across a variety of quantum compilation frameworks.
A single checkmark indicates \emph{compatibility}, i.e., that a framework can consume programs containing the primitive and preserve runtime semantics, even if structure is lost in the process.
A double checkmark indicates that structure is preserved and the compiler is currently engineered to actively transform and optimize these primitives, \emph{exploiting} the embedded information to produce more efficient executables.
While the table shows a positive trend of frameworks beginning to explicitly tackle structured program compilation, the data points clearly show that support remains highly fragmented and incomplete.

An examination of popular formats, such as QIR~\cite{qir} and OpenQASM 3~\cite{cross2022}, reveals a primary cause for this discrepancy.
Fundamentally, neither QIR nor OpenQASM 3 are designed purely as intermediate representations for aggressive compiler transformations; rather, they serve primarily as exchange formats between the compiler stack and the quantum hardware.
Because quantum hardware is historically very restricted in the types of operations it supports natively, wide compatibility relies heavily on straight-line circuits.
Consequently, these formats often err on the side of caution.
They heavily isolate structured elements, for instance by limiting compilers to strictly flat profiles like the QIR Base Profile, or by explicitly marking features like dynamic qubit indexing as optional in OpenQASM 3.
Since the ultimate target architectures often lack native support for structured primitives, there has historically been little incentive for compiler developers to support them.
Due to the ``lowest common denominator'' effect, many toolchains based on these standard exchange formats simply bypass structured control flow altogether.

However, a shift is occurring with IRs that are more deeply rooted in classical compiler infrastructure.
Frameworks building upon MLIR~\cite{mlir, Ittah2024, burgholzer2026a}, as well as quantum-focused IR frameworks like HUGR~\cite{hugr} and Kirin~\cite{kirin}, 
are designed on top of mature classical compilation workflows or take inspiration from them.
As a result, they more naturally incorporate and retain support for structured primitives in optimization passes, pointing toward a more robust handling of advanced control flow natively within the compiler stack.

Addressing the compilation of structured quantum programs therefore boils down to two key aspects: The preservation of structure and its exploitation to produce more efficient executables.
To drive this evolution, it is critical to rely on comprehensive benchmarks.
Benchmarks are the fundamental tool needed to continuously evaluate the state of the art, identify the most pressing gaps in current support, and map the road ahead for fully structured quantum compilation.

\begin{table*}[htbp]
\centering
\caption{List of benchmark programs and their structural features.}
\label{tab:benchmark-algos}
\renewcommand{\arraystretch}{1.06} 
\begin{tabular}{@{} l *{8}{c} @{}}
\toprule
\textbf{Algorithm} & \textbf{StatIter} & \textbf{DynIter} & \textbf{DynVal} & \textbf{CCond} & \textbf{QCond} & \textbf{DynAlloc} & \textbf{DynIdx} & \textbf{Arb. Size} \\ 
\midrule

\multicolumn{9}{@{}l}{\textit{Basic Subroutines \& Primitives}} \\
\quad GHZ (linear \& star)                         & \cmark &        &        &        &        &        & \cmark & \cmark \\
\quad Q. Multiplexers                              & \cmark &        & \cmark & \cmark &        &        & \cmark & \cmark \\
\quad Block Encoding                               &        &        &        &        & \cmark &        &        &        \\
\quad Toffoli-heavy Circuits                       & \cmark &        &        &        &        & \cmark & \cmark & \cmark \\
\quad Parallelization with quantum fan-out         & \cmark &        & $\square$&      &        & \cmark & \cmark & \cmark \\
\quad Teleportation                                &        &        &        &        & \cmark &        &        &        \\
\addlinespace

\multicolumn{9}{@{}l}{\textit{Transforms \& Arithmetic}} \\
\quad QFT                                          & \cmark &        & $\square$&      &        &        & \cmark & \cmark \\
\quad QPE                                          & \cmark &        & \cmark &        &        &        & \cmark & \cmark \\
\quad Iterative QFT                                & \cmark &        & $\square$&      &        &        & \cmark & \cmark \\
\quad Iterative QPE                                & \cmark &        & \cmark &        &        &        & \cmark & \cmark \\
\quad QFT adder (quantum input, two registers)     & \cmark &        & $\square$&      &        &        & \cmark & \cmark \\
\quad QFT adder (classical input, single register) & \cmark &        & \cmark & $\square$&      &        & \cmark & \cmark \\
\quad Controlled multiplication modulo N           & \cmark & \cmark & \cmark & $\square$&      &        & \cmark & \cmark \\
\addlinespace

\multicolumn{9}{@{}l}{\textit{Heuristics, Optimization \& Search}} \\
\quad Grover's Search                              & \cmark &        &        & $\square$&      &        &        & \cmark \\
\quad Grover's Search with Weak Measurement        &        & \cmark &        & $\square$ & \cmark &        &        & \cmark \\
\quad VQE Ansatz with Fixed Repetitions            & \cmark &        & \cmark &        &        &        &        & \cmark \\
\quad QAOA with Fixed Repetitions                  & \cmark &        & \cmark &        &        &        &        &        \\
\quad VQE                                          & \cmark &        & \cmark &           & \cmark &        &        & \cmark \\
\quad ML-QAE                                       &        & \cmark & \cmark      & \cmark         & \cmark &        & \cmark &        \\
\addlinespace

\multicolumn{9}{@{}l}{\textit{Complex Algorithms \& Simulation}} \\
\quad Shor's Algorithm                             & \cmark & $\square$ & \cmark &        &        &        & \cmark & \cmark \\
\quad X-Ray Absorption Spectroscopy                & \cmark &           & \cmark      & \cmark      &        &        & \cmark &        \\
\addlinespace

\multicolumn{9}{@{}l}{\textit{Error Correction \& Fault Tolerance}} \\
\quad Repeat-Until-Success                         &           & \cmark &        &        & \cmark &        &           &        \\
\quad Magic State Distillation                     & \cmark    & \cmark &        &        & \cmark &        & \cmark    & \cmark \\
\quad Logical State Preparation                    & \cmark    & \cmark &        &        & \cmark &        & \cmark    & \cmark \\
\quad Syndrome Measurement and Correction          & \cmark    & \cmark &        &        & \cmark &        & \cmark    & \cmark \\
\quad Measurement-based quantum computation        & $\square$ &        & \cmark &        & \cmark &        & $\square$ &        \\

\bottomrule
\end{tabular} \\ \vspace{0.3em}
{\scriptsize \hfill \cmark: Element required \hfill Blank: Element not required \hfill $\square$: Implementation-dependent \hfill} 
\vspace{-2.2em}
\end{table*}

\section{A Challenge for Structured Quantum Program Compilation}
\label{sec:challenge}
To move the ecosystem beyond its current reliance on straight-line circuits, we propose an open challenge to the developers of quantum compilers.
This challenge introduces a comprehensive benchmark suite of quantum programs that employ varying degrees of structural elements.
We align the evaluation of compilation frameworks with two core requirements: \emph{support} and \emph{utilization}.
First, to establish baseline \emph{support} within state-of-the-art compiler frameworks, representations must be compatible with the individual elements and compilers must handle them effectively.
Second, to drive the \emph{utilization} of these concepts, compilers must be able to natively leverage these elements when compiling structured programs.

\autoref{tab:benchmark-algos} presents the complete benchmark suite, detailing the structural elements required by each program.
These programs were selected to represent a wide range of realistic algorithms covering the different structural elements discussed above.

Some programs mainly feature statically-bounded loops (StatIter), conditionals on classical values (CCond), and other static elements that can be translated to straight-line circuits during preprocessing.
While this makes baseline \emph{support} for these features trivial, compilers must specifically preserve and leverage the provided structure to truly drive \emph{utilization}.

For instance, \emph{Grover's Search} relies heavily on \emph{statically-bounded loops} (StatIter).
The number of iterations can be precomputed, allowing the program to be represented as a fixed number of $N$ iterations over the \emph{oracle} and \emph{diffusion} steps typically employed for the algorithm.
To achieve baseline \emph{support}, a naive quantum compiler can handle this by transforming the program into a flat sequence of $N$ consecutive repetitions.
However, doing so forces the optimizer to independently process all gates each iteration, a process that grows exponentially inefficient since the number of iterations scales as $N = \mathcal{O}(\sqrt{2^n})$ for an $n$-qubit search space.
Instead, by preserving the structure and performing optimizations only within the loop body, an advanced compiler can \emph{utilize} the structure to drastically reduce memory and runtime overhead.

Conversely, other benchmarks feature dynamic elements that intrinsically rely on (quantum) runtime results, making it impossible to flatten them into straight-line circuits.
To sufficiently \emph{support} these algorithms, compilers must explicitly implement native handling for elements such as dynamically-bounded loops (DynIter), conditions on measurements (QCond), or dynamic qubit indexing (DynIdx).

Program classes such as \emph{Repeat-Until-Success} provide a classic baseline for testing these dynamic capabilities.
The contained loops require dynamically-bounded execution (DynIter) to validate that a desired state has been successfully prepared, an operation has been correctly applied, or a specific parity check has passed.
Explicit representations and targeted optimizations are required to correctly parse and process these program classes.
Importantly, successfully utilizing these elements is not solely a compiler problem.
Such advanced control-flow operations require close coordination with the actual control systems of the target quantum hardware~\cite{ransford2025}.
A compiler can only represent and optimize dynamic operations to the extent that the underlying physical devices and their classical control logic can efficiently execute them at runtime.

\smallskip

A benchmark suite is more than a collection of examples for individual structural features; its true value lies in allowing developers to evaluate the scalability of their frameworks.
To facilitate comprehensive scalability studies, many program classes included in this challenge can be scaled to an arbitrary size.
\emph{GHZ preparation}, for instance, can be defined over any number of qubits.
By instantiating these challenge programs at varying sizes, compiler developers can stress-test the compilation times and memory footprints of their pipelines, ensuring their tools remain robust even for massive program instances.

The complete benchmark suite is distributed using the \texttt{jeff} exchange format~\cite{jeff}.
\texttt{jeff} has been specifically created to exchange structured quantum programs between different compilers and frameworks while ensuring native support for all proposed structural elements.
This makes it as easy as possible for developers to get started with the benchmark suite by translating from one common standard to their format of choice.
Conversions from \texttt{jeff} to major open-source SDKs are either already available or currently under development, including PennyLane~\cite{bergholm2018pennylane}~/~Catalyst~\cite{Ittah2024}, HUGR~\cite{hugr}, Kirin~\cite{kirin}, and the MQT Compiler Collection (\texttt{mqt-cc})~\cite{burgholzer2026a}.

\section{Evaluation: case study}

To trace the compatibility of existing compilation frameworks with structured quantum programs, we investigate to what degree the \emph{Repeat-Until-Success} (RUS) benchmark can be processed by state-of-the-art compilers.
To this end, we divide the compilation pipeline into three distinct steps:
\emph{Parsing} programs containing structural elements, \emph{optimizing} them to reduce total resource requirements, and \emph{lowering} them to a format that can be executed on physical quantum hardware.

By investigating the individual compiler API specifications, we come to the following conclusions:
\emph{Qiskit's}~\cite{qiskit} API does not allow the specification of dynamically-bounded \texttt{while} loops.
Consequently, for \emph{Qiskit}, the compilation pipeline of RUS fails already at the \emph{parsing} level.
\texttt{mqt-cc}, on the other hand, explicitly allows the specification of structured programs with dynamically-bounded loops and preserves them during optimization. Whether the lowering step succeeds depends on the targeted output format: when targeting the \emph{QIR Adaptive Profile}, \texttt{mqt-cc} can lower such structured programs to an executable format, whereas the \emph{QIR Base Profile} does not support these constructs and the pipeline fails at the lowering level. For \texttt{mqt-cc}, the compilation pipeline of RUS is therefore complete in principle, but only for a subset of the supported output formats
Going further, the \emph{TKETv2} compiler can compile structured programs in the Guppy language~\cite{koch2025} to HUGR, and lower to  \emph{streaming execution} on the \emph{Helios} quantum computer~\cite{ransford2025}, allowing RUS to follow the full compilation pipeline down to the execution on quantum hardware. Similarly, the Catalyst compiler supports such structure at the language level (via PennyLane), compiler representation and transformations, and execution level --- supporting an array of simulators but no current hardware devices. 

This case study shows that, while some individual compilation frameworks already fully or partially support structured quantum programs, compatibility is far from universal and further effort must be invested in state-of-the-art compilers to make the execution of structured quantum programs possible.

\section{Conclusion and next steps}

In this work, we have motivated the transition from simple straight-line quantum circuits to \emph{structured quantum programs}.
We have defined the individual aspects of structured programs and constructed a taxonomy to facilitate the development of next-generation compilers that are compatible with these more advanced structures.
Incorporating structure into quantum compilers increases their efficiency and allows quantum computing users to develop more scalable and readable programs.

Developing more general, automated higher-level optimization strategies that leverage program structure is an exciting and open area of research.
The challenge programs we provide \cite{jeff} are a call to the community to start incorporating support for structured programs in their quantum compilers, so that future quantum programs can natively exploit these features, much like classical structured control flow has long been a staple in traditional software development.

\section*{Acknowledgments}

D.R., D.H., R.W., and L.B. acknowledge funding from the European Research Council
(ERC) under the European Union’s Horizon 2020 research and innovation
program grant agreement No. 101001318 and No. 101114305 (“MILLENIONSGA1” EU Project), and the Munich Quantum Valley, which is supported by
the Bavarian state government with funds from the Hightech Agenda Bayern
Plus. Furthermore, this work was supported by the BMFTR under grant
numbers 13N17298 (SYNQ) and 01MQ25001I (FullStaQD), the Deutsche
Forschungsgemeinschaft (DFG, German Research Foundation) under grant
numbers 563402549 and 563436708, and the Austrian Research Promotion
Agency (FFG) together with the states of Upper Austria and Tyrol within the
COMET module Quantum Algorithm Engineering (FFG) under grant number
923923.

ODM acknowledges funding from NSERC, the Canada Research Chairs program, and UBC.

\bibliographystyle{IEEEtran}
\bibliography{main.bib}

\end{document}